\documentstyle[preprint,aps,epsfig,subfigure]{revtex}
\newcommand{\be}{\begin{equation}}
\newcommand{\ee}{\end{equation}}
\newcommand{\bea}{\begin{eqnarray}}
\newcommand{\eea}{\end{eqnarray}}
\newcommand{\ba}{\begin{array}}
\newcommand{\ea}{\end{array}}
\newcommand{\aaprime}{\frac{a^{\prime}}{a}}
\newcommand{\ddprime}{\prime \prime}
\begin{document}
\draft
\title{CMBR Anisotropy  with Primordial Magnetic Fields}
\author{Seoktae Koh
\thanks{e-mail: kst@hepth.hanyang.ac.kr} and  Chul H. Lee
\thanks{e-mail: chlee@hepth.hanyang.ac.kr}}
\address{Dept. of Physics, Hanyang University, 
Seoul 133-791, Korea}
\maketitle
\begin{abstract}
Galactic magnetic fields are observed of order $\sim 10^{-6}G$, but
their origin is not definitely known yet.
In this paper we consider the primordial magnetic
fields generated in the early universe and analyse their effects
on the density perturbations and the CMBR anisotropy.  We assume that the
random magnetic fields have the power law spectrum and satisfy the
force-free field condition.
 The peak heights of the CMBR anisotropy are shown to
be shifted upward
depending on the magnetic field strengths
relative to the no-magnetic field case.
\end{abstract}
\newpage
\narrowtext 
\section{Introduction}

Recently many possible generation mechanisms of 
primordial magnetic fields 
have been suggested to explain observed galactic magnetic
fields  of order $\sim 10^{-6}G$ \cite{kronberg}.
The dynamo mechanism explains
the origin of the large scale galactic magnetic field 
with amplification of a small
frozen-in seed field to the observed $\mu G$ field through turbulence 
and differential rotation. The dynamo saturates when the growth enters
the nonlinear regime. However the saturation might actually be too fast
for a large scale field to form \cite{vainshtein}.
Without the dynamo mechanism,
to explain the galactic fields from the primordial fields which get compressed
when the protogalactic cloud collapses, the needed amplitude of the primordial 
magnetic fields is quite large to be of the order of $10^{-10} \sim 10^{-9}G$.
Cosmological phase transitions
in the early universe may  produce magnetic seed fields.
If conformal invariance is
broken during the inflationary period, 
magnetic seed fields are  generated 
\cite{turner}. And 
also the electroweak phase transition 
\cite{baym} \cite{sigl} and QCD phase transition 
\cite{quashnock} \cite {cheng} can generate
 magnetic seed fields. Gasperini et al. \cite{gasperini} considered
generation mechanism in stringy model with broken conformal invariance 
by a dilaton field. 
But the field amplitudes produced by several mechanisms
are much too weak to explain observations.

 Primordial magnetic fields may generate
density perturbations
\cite{peebles} \cite{wasserman} \cite{kim}.  Tsagas and Barrow 
 \cite{tsagas} \cite{tsagas2} 
considered
the general relativistic density perturbations with magnetic fields. To 
treat the large scale cosmological perturbations we confront with the 
gauge ambiguity problem. It is caused by the regions larger than horizon
size being causally disconnected. Bardeen formulated the gauge invariant 
method to solve the gauge ambiguity problem \cite{bardeen}.
The details about
the gauge invariant method of cosmological perturbations 
can be found in Ref's \cite{bardeen}, \cite{kodama} and \cite{mukhanov}.
Cosmological perturbations can be classified according
to how they transform under spatial coordinate transformations in the
background spacetime; scalar, vector, and tensor perturbations. 
They relate
to density, vorticity and gravitational wave perturbations respectively.
  Here in this 
paper we only consider scalar perturbations.
 From the observations that the magnetic
field energy density is much less than the background radiation energy 
density, we can treat it within the linear perturbation theory.

The Big Bang Nucleosynthesis(BBN) can constrain the amplitude of
magnetic fields at early epochs. It is argued in Ref. \cite{cheng2}
that the  presence
of magnetic fields affects BBN by changing the weak reaction rates, 
the electron density and  the expansion rate of 
the universe. So they put constraint on the magnetic field amplitude
 $B < 2 \times 10^9 G$ at  $T= 0.01Mev$. Barrow et al. \cite{barrow}
also derived an upper limit of the magnetic field amplitude at present
 $B_{0}
< 3.4 \times 10^{-9} G$  using microwave background anisotropy created 
by cosmological magnetic fields.

The CMB photons are polarized through the Thomson scattering of 
the photons and  electrons during the decoupling time \cite{chandra}.
 The upper limit on
its degree of linear polarization large angular scales is 
$\Delta_P < 6 \times 
10^{-5}$ \cite{lubin}. 
We expect that the CMBR polarization on 
small angular scales would be observed with the  future experiments, 
MAP \cite{map} and PLANCK \cite{planck}.
 If  primordial 
magnetic fields exist at the decoupling time, they cause Faraday
rotations which rotate the directions of polarization
vectors. This effect can be imprinted on the cosmic background radiation
and  we may obtain informations on
 the amplitude of primordial magnetic fields
by measuring  the polarizations  of the CMBR \cite{kosowsky}
\cite{harari}. 

In this paper  we calculate the evolution of density perturbations with the
primordial  magnetic 
fields which have power law spectrum. We do not concern ourselves with
 the details
of  generation mechanism of magnetic seed fields, but assume  that 
sometime  during
 radiation dominated era, 
large scale magnetic fields are generated instantaneously. We then 
investigate how they affect 
the temperature anisotropy and polarization 
of the CMBR using various  spectral
indices and field strengths of the magnetic field.

In section \ref{sect}, we derive, 
using the gauge invariant variable, the
density 
 perturbation equations with magnetic fields present.
 The  equations are solved numerically and the effect of magnetic fields on 
the temperature anisotropy and polarizations of the CMBR 
 are shown in section III. 
 Finally we summarize the results
 in section IV.

\section{Density perturbations with Magnetic Fields} \label{sect}
In this section we consider the 
background space is homogeneous and
isotropic. Cosmological perturbations are  classified as scalar, vector 
and tensor perturbations according to 
 how they  transform under  spatial coordinate
transformations in the background spacetime. 
We will treat here only scalar perturbations which 
are related to density perturbations. 
In the longitudinal gauge (conformal Newtonian gauge) 
the metric, including the scalar perturbations, is written by
\cite{mukhanov} ,
\bea
ds^{2}=a(\eta)^{2}\left(-(1+2 \Psi) d \eta^{2} + (1+2 \Phi) 
\gamma_{ij} dx^{i} dx^{j} \right)
\eea
where $\eta$ is the conformal time defined by $dt = a(\eta) d
\eta$ and $\gamma_{ij}$ is the spatial metric tensor.
 $\Psi$ and $\Phi$ are related to the gauge invariant quantities
$\Phi_{A}$ and $\Phi_{H}$ of Bardeen \cite{bardeen}  and the gauge
 invariant potentials $\Psi$ and $\Phi$  of
Kodama $\&$ Sasaki \cite{kodama}.
 The physical meaning
of $\Phi$ and $\Psi$ are the curvature perturbation and Newtonian
gravitational potential respectively.

The Maxwell equations have the form
\begin{eqnarray}
& & {F^{\mu \nu}}_{; \nu}=  J^{\mu}  \\
& & F_{\mu \nu ;\rho}+F_{\nu \rho; \mu}+F_{\rho \mu; \nu} = 0,  \label{maxwell}
\end{eqnarray}
where $F_{\mu \nu}$ is the second-order antisymmetric Maxwell tensor
which represent the electromagnetic field and $J_{\mu}$ is the
4-vector current which generates the electromagnetic field.
The Maxwell tensor  splits into the electric and magnetic $4$-vector
 \cite{tsagas}, defined by,
\bea
& & E_{\mu}= F_{\mu \nu} u^{\nu},  \\
& & B_{\mu}=\frac{1}{2} \epsilon_{\mu \nu \rho \lambda} u^{\nu} F^{\rho \lambda} ,
\eea
where $\epsilon_{\mu \nu \rho \lambda}$ is Levi-Civita tensor and $u^{\mu}$ is 
the fluid four-velocity. 
The background value of $u^{\mu}$ is taken to be $u^{\mu}=(1/a,
0, 0, 0)$. Then the electric and magnetic 
4-vectors are purely spatial, i.e. $E^{\mu} u_{\mu}
=0$ and $B^{\mu} u_{\mu} =0$, so we denote the spatial components
$E^{i}$ and $B^{i}$ by 
 ${\bf E}$
and ${\bf B}$.

The generalized covariant Ohm's law is
\bea
J^{\mu} + J^{\nu} u_{\nu} u^{\mu} = \sigma F^{\mu \nu} u_{\nu} \label{ohm}
\eea
where $\sigma$ represents  conductivity of the medium.
The  spatial components of Eq. (\ref{ohm}) are reduced to 
${\bf J}= \sigma {\bf E}$ where ${\bf J}$ is the spatial component of
$J^{\mu}$.  
 Assuming infinite conductivity of the medium 
 in the Universe \cite{turner}, 
we neglect the electric field so that ${\bf E} =0$.

Now we can reduce Eq. (\ref{maxwell}), 
using the magnetic field 3-vector ${\bf B}(\eta,{\bf x})$, to
\bea
\frac{\partial(a^{3} {\bf B})}{\partial \eta} = 0,~~~~ 
\nabla \cdot {\bf B}=0,  \label{max2}
\eea
where $\nabla$ is the covariant differentiation with respect to $\gamma_{ij}$.
In this work, we consider only the case where $\gamma_{ij}=\delta_{ij}$
for simplicity.
From the first of Eq. (\ref{max2}), we 
 get   ${\bf B}(\eta, {\bf x}) \propto a^{-3}$. 
The magnetic field energy density, $\frac{1}{2} B^{2} 
(=\frac{1}{2} B^{\mu}B_{\mu}=\frac{1}{2}a^2
\sum_{i=1}^{3}B^{i}B^{i})$, evolves the same as the
radiation energy density  $\sim a^{-4}$. 
The dimensionless
quantity $r$ is introduced  in Ref. \cite{turner} defined by  
$r= \frac{B^{2}}{2 \rho_{r}}$  which is the ratio of magnetic
field energy density to the background  radiation energy density. It
is  approximately constant at all early history of the Universe.  

Total energy momentum tensor is  decomposed by
\bea
T^{\mu \nu} = T^{(fluid) \mu \nu} + T^{(em) \mu \nu} ,
\eea
where the electromagnetic energy momentum tensor  and fluid energy
momentum tensor are given by
\bea
T^{(em)\mu \nu} = {F^{\mu}}_{\lambda} F^{\nu \lambda}- \frac{1}{4}
g^{\mu \nu} F_{\rho \sigma} F^{\rho \sigma} ,
\eea
and
\bea
T^{(fluid)\mu \nu} = (\rho + P)u^{\mu} u^{\nu} + Pg^{\mu \nu} .
\eea
We here treat the matter as perfect fluid to neglect the anisotropic
pressure perturbations and consider only adiabatic perturbations to
neglect the entropy perturbations.
The linearized perturbation equations are 
obtained  from the Einstein equations up to first 
order, and are written as follows;
\bea
& & 3 \left(\aaprime\right)^{2} \Psi - 3 \aaprime \Phi^{\prime} +(\nabla^{2} -3K) \Phi = -4
\pi G a^{2} \rho\left(\Delta + \frac{1}{2}\frac{B^{2}}{\rho}\right), 
\label{eq11} \\
& & \nabla_{i} (\aaprime \Psi - \Phi^{\prime}) = -4 \pi G a^{2} (\rho+P) v_i  \\
& & \Phi^{\ddprime} +\aaprime (2 \Phi^{\prime} -\Psi^{\prime})+ 
\left(\left(\aaprime\right)^{2}
 -2 \frac{a^{\ddprime}}{a}\right)\Psi + \frac{1}{3}\nabla^{2} \Psi
-\frac{1}{3} (\nabla^{2} + 3 K) \Phi   \nonumber  \\
& &~~~~~~ = - 4 \pi G a^{2}
 \rho \left(c_{s}^{2} \Delta  + \frac{1}{6} \frac{B^{2}}{\rho}\right)  \label{trace} \\  
& & \left(\nabla^{i}\nabla_{j} - \frac{1}{3}\delta^{i}_{j} \nabla^{2}\right)
 (\Phi + \Psi)= -8 \pi G a^2 {\Pi^{(em) i}}_{j}  \label{treless}
\eea
where the prime denotes derivative with respect to the conformal time
$\eta$, and ${\Pi^{(em)i}}_{j}= \frac{1}{3}\delta^{i}_{j}B^{2}-B^{i}B_{j}$ 
corresponds to the magnetic field anisotropic pressure.
 $\Delta$ and $ v_{i}$ are the gauge invariant density contrast
 and  velocity perturbation and $c_{s}$ is the sound velocity. 
 $\nabla^{2}$ is the  Laplace-Beltrami operator whose
 eigenvalue is $-k^2$.
Eq. (\ref{trace}) is the trace part of the 
spatial component of the perturbed energy-momentum 
tensor  and Eq. (\ref{treless})
is the traceless part.

To write down the perturbation equations for a given wave mode, ${\bf k}$,
we  define the Fourier transform of the perturbed quantities and random 
magnetic fields. In this paper we consider density perturbations in  
flat space, $K=0$, so the spatial
dependence of the Fourier transform is just  
the plane wave, $e^{i {\bf k}\cdot {\bf x}}(=e^{i k_{i}x^{i}})$,
\bea
\Delta({\bf x})=\int d^{3}k~~ {\rm exp}(i {\bf k} \cdot {\bf x}) 
~~\Delta({\bf k}) 
\eea
and also $v({\bf x})$, $\Phi({\bf x}), \Psi({\bf x})$ and $B({\bf x})$  
are defined similarly.
We assume that the force-free field condition
($ {\bf B} \times \nabla \times {\bf B} = 0$) is satisfied  
to treat magnetic field. Then Eq.'s (\ref{eq11}) $\sim$ (\ref{treless})
can be written by
\bea
& & 3 (\aaprime)^{2} \Psi - 3 \aaprime \Phi^{\prime} -k^2 \Phi = -4
\pi G a^{2} \rho\left(\Delta + \frac{1}{2}\frac{ F(k)}{\rho a^{-4}}\right), \label{energy} \\
& & k (\aaprime \Psi - \Phi^{\prime}) = -4 \pi G a^{2} (\rho+P) v  \\
& & \Phi^{\ddprime} +\aaprime (2 \Phi^{\prime} -\Psi^{\prime})+ \left((\aaprime)^{2}
 -2 \frac{a^{\ddprime}}{a}\right)\Psi + \frac{k^2}{3} \Psi
+\frac{1}{3} k^2  \Phi   \nonumber \\
& &~~~~~~ = - 4 \pi G a^{2}
 \rho \left(c_{s}^{2} \Delta  + \frac{1}{6} \frac{ F(k)}{\rho a^{-4}}\right)  \\  
& & k^2(\Phi +\Psi)= - 8 \pi G a^{-2} \left(\frac{1}{4}  F(k)\right)  \label{tless}
\eea
where $ F(k)$ is defined by
\bea
 F(k) = \int d^{3}k^{\prime} B^{l}({\bf k^{\prime}})B_{l}({\bf k-k^{\prime}}). \label{fk}
\eea
which represents the spectral dependence of magnetic fields. 
The fact that the magnetic
field energy density decays as $ \sim a^{-4}$ is used. In the Appendix A,
 we calculate the 
Fourier transform of the magnetic field anisotropic pressure
using force-free field conditions
and derived the expression $F(k)$,Eq. (\ref{fk}).

To investigate the spectral dependence of perturbed quantities,
 we need to take ensemble
average of $|F(k)|^2$ due to random magnetic field.  
 For a homogeneous and isotropic random
magnetic field, ${\bf B}({\bf k})$ satisfy the relation \cite{wasserman}
\cite{kraichnan}, 
\bea
< B^{i}({\bf k}) B^{j *}({\bf k^{\prime}})>   
 = \delta^{3} ({\bf k} -
   {\bf k^{\prime}}) \left(\delta_{ij} - \frac{k_{i} k_{j}}{k^{2}}\right) \frac{B^{2}(k)}{2},
 \label{ensemble}
\eea
and then
\bea
<B_{0}^{2}> = \int d^{3}k B^{2}(k)   \label{present}
\eea
where angular brackets denote a statistical  average over an ensemble
of possible magnetic field configurations and 
$<B_{0}^{2}>^{1/2}$ is the average field strength observed
today.

From Eq. (\ref{fk})
\bea
| F(k)|^2 =\int d^{3}k^{\prime}~ d^{3}k^{\ddprime} ~{\bf B({\bf k^{\prime})}}
\cdot {\bf B({\bf k-k^{\prime}})} {\bf B({\bf k^{\ddprime}})}^{*} \cdot
{\bf B({\bf k-k^{\ddprime}})}^{*} 
\eea

Taking ensemble average of the both sides, we obtain
\bea
<| F(k)|^2> & &= \int d^{3}k^{\prime}~ d^{3}k^{\ddprime}
  <{\bf B({\bf k^{\prime})}}
\cdot {\bf B({\bf k-k^{\prime}})} {\bf B({\bf k^{\ddprime}})}^{*} \cdot
{\bf B({\bf k-k^{\ddprime}})}^{*}>   \nonumber  \\
& &= \int d^{3}k^{\prime}~ d^{3}k^{\ddprime}~ [<B^{l}({\bf k-k^{\prime}})
B^{* m}({\bf k- k^{\ddprime}})><B_{l}({\bf k^{\prime}})B^{*}_{m}({\bf k^{\ddprime}})> \nonumber \\
& &+ <B^{l}({\bf k-k^{\prime}})B^{*}_{m}({\bf k^{\ddprime}})>
<B_{l}({\bf k^{\prime}})B^{* m}({\bf k-k^{\ddprime}})>]
\eea
Using Eq. (\ref{ensemble}), and integrating the delta function, we can get
\bea
<|F(k)|^2> &=& 2 \int d^{3}k^{\prime} 
\left[ 1 + \frac{\{({\bf k-k^{\prime}}) \cdot 
{\bf k^{\prime}}\}^2}{|{\bf k-k^{\prime}}|^{2} k^{\prime 2}}\right]
\frac{B^{2}(|{\bf k-k^{\prime}}|)}{2} 
\frac{B^{2}(k^{\prime})}{2}  \nonumber  \\
&=& \frac{A^2 V}{8 \pi^2} \int^{k_{max}}_{0}  
dk^{\prime} k^{\prime q+2} \int^{1}_{-1}
d\mu [k^{2}(1+\mu^{2})+ 2k^{\prime 2} - 4 k k^{\prime}\mu] \nonumber  \\
& & \times [k^2 + k^{\prime 2}-2 k k^{\prime} \mu]^{\frac{q}{2}-1} 
 \label{int}
\eea
where  ${\bf k}\cdot {\bf k^{\prime}}= k k^{\prime} 
\mu$, $\mu={\rm cos} \theta$ and  $\delta^{3}(k=0)
=V/(2 \pi)^3$ is used.
$V$ is the volume factor.
We used the power-law spectrum
\bea
B^{2}(k)= A k^{q}.  \label{power}
\eea
We define the average field on scale $\lambda$ by
\bea
<B^{2}>_{\lambda} =\int d^3 k B^{2}(k) 
{\rm exp}\left(\frac{-k^{2}{\lambda}^{2}}{2}\right).
\eea
Then we can determine the coefficient $A$ in Eq. (\ref{power})
\bea
A = \frac{1}{4 \pi}\frac{\lambda^{q+3}}
{2^{(n+1)/2}\Gamma(n+\frac{3}{2})}<B^2>_{\lambda} .
\eea
To determine $k_{max}$ we use the argument in Ref. \cite{kim}. 
Small scales reach nonlinear variance ($\Delta \ge 1$) earlier than 
large scales, and the first scales to become nonlinear have 
$k \approx 2k_{max}$. If we choose the scale that corresponds to the 
formation of galaxies at $z_{nl}=6$, $k_{max}=\pi Mpc^{-1}$. 
If instead we
require that magnetic fields form clusters of galaxies at $z_{nl}=1$, that
would correspond to $k_{max}=\frac{\pi}{2} Mpc^{-1}$.
As we shall see below, the CMBR anisotropy much depends on the $k_{max}$ value.
Treating $k/k_{max}$ as a small parameter, the leading term of the
result of integrating Eq.(\ref{int}) is 
\bea
<| F(k)|^2>  \simeq \frac{\pi V}{(2 \pi)^3}  A^2 \frac{4}{2q +3} 
k_{max}^{2q+3}. 
\eea
For the case of $q \le -3$, the integration of Eq. (\ref{int})
diverge as $k^{\prime} \rightarrow 0$. So we only consider 
$q > -3$.

\section{CMBR anisotropy}
Cosmological density perturbations cause the temperature 
fluctuations when the photons 
decouple from the thermal bath at last scattering surface. 
Furthermore small metric perturbations induce bulk velocities
of the fluid, and the resulting anisotropies in the photon
distribution will induce polarization when the photons scatter
off charged particles (Thomson scattering) \cite{kosowsky}.
After decoupling, the photons freely
propagate along the geodesics, and any polarization produced through
the Thomson scattering will remain fixed.
The evolution of the CMB anisotropies is described by a set of
radiative transfer equations. 
Temperature and polarization anisotropies are expressed in terms of
Stokes  parameters
$I, Q, U$ and $V$. The parameter $I$ gives the radiation intensity  which is
positive definite, $Q$ and $U$ represent the linearly polarized light and
$V$ describes the circular polarization. The degree of linear
polarization $\Delta_{P}$ is defined in terms of $Q$ and $U$,
$\Delta_{P} = (Q^{2} + U^{2})^{1/2}$, and the temperature anisotropy
is denoted by $\Delta_{T} (\equiv \frac{\Delta T}{T})$.
 $\Delta_{T}$ and $\Delta_{P}$ 
 can be expanded in
multipole moments defined such that $\Delta(\eta, k, \mu) = {\sum_{l}} 
(2l+1) \Delta_{l}(\eta, k) P_{l}(\mu)$ where  $P_{l}$ is 
the Legendre polynomial
of order $l$ and $\mu$ is the cosine angle between the wave vector and
the direction of observation.
Their evolution equations are given by\cite{zaldarriaga2}
\bea
& & \Delta^{\prime}_{T}+ i k \mu (\Delta_{T} + \Psi) = - \Phi^{\prime} -
\kappa^{\prime} [ \Delta_{T} - \Delta_{T_{0}} - \mu v_{b} +
\frac{1}{2} P_{2}(\mu) S_{p}], \label{temp} \\
& & \Delta^{\prime}_{P} + i k \mu \Delta_{P} = - \kappa^{\prime}
[\Delta_{P} - \frac{1}{2} (1- P_{2} (\mu)) S_{P} ] \label{pol}
\eea
where  $S_{P} \equiv - \Delta_{T_{2}} - 
\Delta_{P_{2}} + \Delta_{P_{0}}$.
 $\kappa^{\prime}$ is the 
 differential optical depth defined by
$\kappa^{\prime} = x_{e}n_{e}\sigma_{T} a/a_{0}$ with $x_{e}$ the
ionization fraction, $n_{e}$ the electron number density and
$\sigma_{T}$ the Thomson scattering cross section.
The Thomson scattering cannot produce any net circular
polarization \cite{chandra} and  thus we expect $V=0$ for the microwave
background. 
$v_{b}$ is the velocity perturbations of the baryon component 
which is affected by the existence of 
primordial random  magnetic fields. Metric perturbations
$\Phi$ and $\Psi$, which evolve according to the  equations 
in section \ref{sect} under the influence of random magnetic fields,
act as the source terms in Eq. (\ref{temp}) which
governs the evolution of the temperature anisotropy.

Eq.'s (\ref{temp}) and (\ref{pol}) are formally integrated 
to yield \cite{zaldarriaga2}
\bea
& & \Delta_{T}(\eta_{0}) = \int_{0}^{\eta_{0}} d \eta e^{ix \mu}
g(\eta) [\Delta_{T_{0}}(\eta) + \mu v_{b}(\eta) - \frac{1}{2}
P_{2}(\mu) S_{P}(\eta)] + \int_{0}^{\eta_{0}} d \eta e^{ix\mu} e^{-
  \kappa(\eta_{0} ,\eta)}(\Psi^{\prime} - \Phi^{\prime}), \label{delt} \\
& & \Delta_{P}(\eta_{0}) = \int_{0}^{\eta_{0}} d \eta e^{ix \mu}
g(\eta) \frac{1}{2} [1- P_{2}(\mu)] S_{P}(\eta) \label{delp},
\eea
where
\bea
g(\eta) \equiv \kappa^{\prime} e^{- \kappa(\eta_{0}, \eta)}
\eea
is the visibility function and 
\bea
\kappa(\eta_{0}, \eta) = \int^{\eta_{0}}_{\eta} \kappa^{\prime} d\eta
\eea
is the optical depth to photons emitted at the conformal time $\eta$.
 The visibility function  represents the probability that a
photon observed at $\eta_{0}$ last scattered within $d \eta$ of a given
$\eta$. For the standard recombination this function has a sharp peak at
the conformal time of decoupling $\eta_{D}$. And $x= k(\eta_{0}-\eta)$.

Under a clockwise rotation in the plane perpendicular 
to the direction of
observation, $\hat{\bf n}$, the temperature is invariant while
$Q$ and $U$ transform as
\bea
Q^{\prime} = Q {\rm cos}2 \psi + U {\rm sin} 2 \psi,  \nonumber \\
U^{\prime} = - Q {\rm sin}2 \psi + U {\rm cos} 2 \psi,
\eea
or
\bea
(Q \pm i U)^{\prime} = e^{\mp 2i \psi}(Q \pm iU)
\eea 
where $\psi$ is the rotation angle.
Therefore the quantities can be expanded in terms of the  spin-2 spherical
harmonics \cite{zaldarriaga}
\bea
(Q \pm iU)(\hat{\bf n}) = \sum_{lm} a_{\pm 2,lm}
{_{\pm2}}Y_{lm}(\hat{\bf n})
\eea
where $_{\pm 2}Y_{l}^{m}(\hat{\bf n})$ is the spin-2 spherical harmonics
whose  properites are summarzied briefly in Appendix B.
The expansion coefficients are 
\bea
a_{\pm 2,lm}=\int d\Omega {_{\pm2}}Y_{lm}^{*}(Q \pm iU)(\hat{\bf n}).
\eea
In Ref. \cite{zaldarriaga}, the authors introduce the following 
linear combinations of 
$a_{\pm 2,lm}$ to circumvent the difficulty that the Stokes 
parameter are not invariant under rotations;
\bea
a_{E,lm}=-\frac{1}{2}(a_{2,lm}+a_{-2,lm})  \nonumber \\
a_{B,lm}= \frac{i}{2}(a_{2,lm}-a_{-2,lm}).
\eea
These newly defined variables are expanded in terms of ordinary spherical
harmonics, $Y_{lm}$,
\bea
E(\hat{\bf n})= \sum_{lm} a_{E,lm} Y_{lm}(\hat{\bf n}), \nonumber \\
B(\hat{\bf n})= \sum_{lm} a_{B,lm} Y_{lm}(\hat{\bf n}).
\eea
The spin-zero spherical harmonics, $Y_{lm}$, is free from
 the ambiguity with the rotation of the coordinate system,
and therefore $E$ and $B$ are rotationally invariant quantities.
The  $E$-mode  has $(-1)^{l}$ parity and the  $B$
mode $(-1)^{(l+1)}$ parity in analogy with electric and magnetic
fields.
Scalar perturbations generate only the $E$ mode of the polarizations
\cite{kamionkowski}.
The power spectra of temperature and polarization anisotropies
are defined as $C_{Tl} \equiv <|a_{T,lm}|^2>$ for $\Delta_{T} =
\sum_{lm}a_{T,lm}Y_{lm}$ and analogously for $C_{El}$.
So if we get the evolution of the temperature and polarization
anisotropy amplitude from Eq. (\ref{delt}) and (\ref{delp}),
the amplitudes for each mode of power spectra are given by
\bea
& & C_{T, l}= (4 \pi)^{2} \int k^{2} dk P_{\delta} (k)
[\Delta_{T,l}(k)]^{2} ,  \\
& & C_{E, l}= (4 \pi)^{2} \int k^{2} dk P_{\delta} (k)
[\Delta_{E,l}(k)]^{2} ,  \\
& & C_{Cl} = (4 \pi)^{2} \int k^{2} dk P_{\delta} (k) \Delta_{Tl}(k)
\Delta_{El} (k)
\eea
where $P_{\delta}(k)$ is the initial power spectrum 
and $\Delta_{Tl}$ and $\Delta_{El}$ are given by \cite{zaldarriaga},
\bea
& & \Delta_{Tl}(k) = \int_{0}^{\eta_{0}} d \eta S_{T}(k, \eta)
j_{l}(x), \\
& & \Delta_{El}(k) = \sqrt{\frac{(l+2)!}{(l-2)!}}  \int_{0}^{\eta_{0}}
 d \eta S_{E}(k, \eta) j_{l}(x), \\
& & S_{T}(k,\eta)= g(\Delta_{T0}+ \Psi - \frac{v_{b}^{\prime}}{k} -
\frac{S_{P}}{4} - \frac{3 S_{P}^{\ddprime}}{4 k^{2}}) + 
e^{-\kappa} (\Phi^{\prime} +
  \Psi^{\prime}) - g^{\prime} (\frac{v_{b}}{k}+ 
\frac{3 S_{P}^{\prime}}{4 k^{2}})
  - \frac{3 g^{\ddprime} S_{P}}{4 k^{2}} ,  \\
& & S_{E} (k, \eta) = \frac{3 g S_{P}}{4  x^{2}}.
\eea

We here concern ourselves with the flat CDM universe with adiabatic 
initial conditions. We
use the CMBFAST code \cite{seljak} to calculate numerically 
the CMBR anisotropy. 
During this calculations we put $h$ (Hubble 
constant divided by $100 km/sec/Mpc$) $=0.5$
and assume $3$ species of massless neutrinos.
In Fig. 1, we plot the angular power spectrum of temperature 
fluctuations  $l(l+1)C_{Tl}$ with 
the magnetic field strengths, $3 \times 10^{-8}G$, $5 \times 10^{-8}G$ 
and $7 \times 10^{-8}G$, 
for a given magnetic field spectrum index, $q=1$.
Observed amplitude of galactic magnetic fields is order of 
$\sim 10^{-6}G$.
The BBN can constrain the amplitude of magnetic fields, $B_{0}< 10^{-7}G$ 
\cite{cheng2}, and also derived an upper limit
of the magnetic field amplitude $B_{0} < 10^{-9}G$ using
the CMB anisotropy \cite{barrow}.
Another constraint on magnetic field intensity can be obtained from 
$r~(\equiv B^{2}/2 \rho_{r}) \leq \Delta_{H}$ where $\Delta_{H}$ is 
the horizon crossing amplitude. COBE $4$-yr data gives
$\Delta_{H} \sim 10^{-5}$ on large angular scales. 
To consider magnetic fields of order of $\sim 10^{-8}G$ 
 does not violate too much current bounds on magnetic 
field amplitudes by the  observational and 
theoretical considerations.
The figure shows that 
the spectral curves of the CMBR temperature anisotropy
are shifted upward
with increasing magnetic field strengths. We can conclude from the
numerical calculations that the presence of the magnetic fields which have 
field strength of order $\le 10^{-9}G$ at present doesn't 
significantly affect
the temperature fluctuations. 
The density perturbations with magnetic fields of order of $10^{-8}G$
result in the deviations of the angular power spectrum, $C_{Tl}$, of
up to $14\%$.
 
The $E$-polarization spectrum, $l(l+1)C_{El}$, and 
temperature-polarization correlation
spectrum, $l(l+1)C_{Cl}$, are shown in Fig.'s 2 and 3 for $q=1$. 
Also in these figures
we can see that the spectrum curves are shifted upward  
with increasing  magnetic field strengths relative 
to the non-magnetic case. 
The current bound on the degrees of linear polarizations of the 
CMBR on large angular scales is 
$\Delta_{P} < 6 \times 10^{-5}$ \cite{lubin}.
As we discussed in the previous
section, we plot the temperature anisotropy with 
$k_{max}=\pi/2 Mpc^{-1}$ and $\pi Mpc^{-1}$ in Fig. 4.
In this figure, we can see that there is a strong dependence 
of the spectrum curves on the cutoff $k_{max}$. 

 In Fig. 5 we plot the temperature anisotropy with the spectral 
index of magnetic field $q=1,2$ and $3$ for
$B_{\lambda}=5 \times 10^{-8}G$ with $\lambda=0.1h^{-1}{\rm Mpc}$. 
The spectrum curves are nearly
independent of the spectral index. We probe the vicinity of 
the acoustic oscillation peak, $l \simeq
200$, to investigate the dependence of spectral index more closely.
The result is  that the spectrum curves are shifted downward 
with the increasing spectral index. 
In Ref. \cite{durrer} recently, the authors derive
an expression for the angular power spectrum of CMBR anisotropies 
due to gravity waves generated by a stochastic magnetic field. 
They show that, for
$n > -3/2$ ($n$ is magnetic field 
spectral index in their notations), the
induced $C_l$ spectrum from gravity waves is independent of $n$,
but only the amplitude depends on the spectral index, $l^2 C_l 
\sim (\lambda k_{max})^{2n+3} l^3$. They also derive an upper
bound of $B_{\lambda}$ for $n> -3/2$ and $\lambda =0.1h^{-1}$Mpc
\bea
B_{\lambda} ~~ <~ 9.5 \times 10^{-8} e^{-0.37n} G.
\eea

Here we don't consider the Faraday rotation due 
to the magnetic field which can 
change the polarization spectrum because we 
restrict our calculations in linear
perturbation theory.
The authors in Ref's \cite{kosowsky}, 
\cite{harari} and \cite{scannapieco}
studied the effect on the CMBR anisotropy 
with the uniform primordial magnetic
field  causing Faraday rotations in the homogeneous background universe.
  They argued that the presence of
magnetic fields depolarize the CMBR anisotropy \cite{harari} and
proposed that the temperature and $B$ mode polarization correlation 
 which are generated by Faraday rotations  can constrain the
magnetic field \cite{scannapieco}.

\section{Conclusions and Discussions}
In this paper we consider the density 
perturbations with primordial magnetic
fields present using gauge invariant formalism. While magnetic field
generation mechanism is not yet known, we assume that the magnetic fields
smear out all over the universe randomly in the radiation dominated
era. Using the CMBFAST code \cite{seljak} we solve numerically the
coupled  density perturbation equations for flat
 CDM universe with adiabatic initial conditions.
 We investigate the CMBR anisotropies for the magnetic field  permeated
universe. With the temperature anisotropy 
spectrum we cannot fully determine
the cosmological parameters and the informations about
the cosmological perturbations. 
CMB photons are polarized due to Thomson scattering during the
decoupling time. Small cosmological perturbations induce the
polarization at last scattering surface through the Thomson
scattering. The linear polarization relates to the quadrupole
anisotropy in the photons. So if we investigate the polarization as
well as the temperature anisotropy, we can get enough information at
the time of decoupling. The information can constrain the   
cosmological parameters and  the cosmological perturbations.
 Next  we study  the effect of the  random magnetic field on the
 CMBR temperature anisotropies and polarizations. We consider the several
 scale magnetic fields with the assumption of power law magnetic spectrum. 
 To get the polarization power spectra, we use
the rotation invariant scalar quantities $E$ and $B$
which are introduced in Ref. \cite{zaldarriaga}. $B$  vanishes for
scalar perturbations. 

For a  given spectral
index  the temperature anisotropy and polarization spectrum are
shifted upward  with
increasing  magnetic field strengths. 
The density perturbations with magnetic fields of order of
$10^{-8}G$ result in the deviations of temperature anisotropy
power spectrum of up to $14\%$.
 The fluctuations of this order due to the primordial
magnetic field  are sufficiently large to be  observed 
in future satellite
experiments. Further, if $B_{\lambda} < \sim 10^{-9}G$, 
magnetic field 
energy density does not affect noticeably the CMBR anisotropy. 
We assume that the magnetic fields have the power-law spectrum.
The spectrum curves are nearly independent of the magnetic
field spectral index. 

Here we assume that the magnetic field energy
density evolves as $\sim a^{-4}$.
But in the
early era, when the magnetic fields are generated, their evolution 
behaviors may be different depending on generation mechanism. If so, 
temperature fluctuations due to magnetic field may be 
shown.

In the early next century, the new satellite experiments, 
MAP \cite{map} and
PLANCK \cite{planck}, will be set forth with better accuracy 
than COBE satellite. They are
 expected to detect the imprint of the polarization as well as
gravitational wave. If it is possible, we can constrain the magnetic
field strength  and the spectral index and get the hint about the 
magnetic field generation mechanism.   

\acknowledgments
CHL is supported by Hanyang University. This work was also 
supported by grant No. 1999-2-112-003-5 from the interdisciplinary
Research program of the KOSEF.

\appendix
\section{Calculation of traceless part of pressure perturbation}
Here we calculate the Fourier transform 
of random magnetic fields with the assumption of 
for the force-free
magnetic field (${\bf B} \times \nabla \times {\bf B}=0$) condition.
From the perturbed Einstein equations, 
we can write the traceless part of pressure
perturbations as follows in the real space,
\bea
(\nabla^i \nabla_j - \frac{1}{3}\delta^{i}_{j})(\Phi + \Psi) = 
- 8 \pi G a^2 {\Pi^{(em) i}}_{j}, \label{atrace}
\eea
where ${\Pi^{(em)i}}_{j}=\frac{1}{3}\delta^{i}_{j}B^{2}-B^{i}B_{j}$.
Using the Fourier transform of ${\bf B}({\bf x})$,
${\Pi^{(em) i}}_{j}({\bf x})$ can be written by
\bea
{\Pi^{(em)i}}_{j}({\bf x})= \int d^{3}k d^{3}k^{\prime} 
[\frac{1}{3}\delta^{i}_{j}
B^{l}({\bf k})B_{l}({\bf k-k^{\prime}}) - B^{i}({\bf k})B_{j}({\bf k-k^{\prime}})]
e^{i{\bf k \cdot x}}
\eea
where we omit the time dependence for brevity.
Then we differentiate ${\Pi^{(em) i}}_{j}({\bf x})$ 
with respect to $x^i$ to get,
\bea
\nabla_{i} {\Pi^{(em) i}}_{j}({\bf x})= 
\int d^{3}k d^{3}k^{\prime} ik_{i} [\frac{1}{3}\delta^{i}_{j}
B^{l}({\bf k})B_{l}({\bf k-k^{\prime}}) - B^{i}({\bf k})B_{j}({\bf k-k^{\prime}})]
e^{i{\bf k \cdot x}}   \label{fourier}
\eea
We can assume
\bea
\int d^{3} k^{\prime} B^{i}({\bf k})B_{j}({\bf k- k^{\prime}}) 
=(A\delta^{i}_{j}+
B\frac{k^{i}k_{j}}{k^2})k^2 F(k)
\eea
which is obvious from the fact that tensorial component
 of scalar perturbations
 is split into the trace  and traceless part.

Next, differentiating $B^{i}B_{j}$ with respect $x^i$ yields
\bea
\nabla_{i} (B^{i}B_{j})= ({\bf B}\times \nabla \times {\bf B})_{j} 
+ B^{i}\nabla_{j}B_{i}.
\eea
The first part of the right-hand side is the 
magnetic force due to the current density(${\bf J}
= \nabla \times {\bf B}$), 
and we neglect this term assuming that the force-free
condition is satisfied in the early universe.
We take Fourier transform of the ${\Pi^{(em) i}}_{j}$  
and then again differentiate with respect to $x^i$
using the force-free field condition to obtain,
\bea
\nabla_{i} {\Pi^{(em) i}}_{j}({\bf x}) = \int d^{3}k \int d^{3}k^{\prime} ik_{j}[\frac{1}{3}
B^{l}({\bf k^{\prime}})B_{l}({\bf k}-{\bf k^{\prime}}) - \frac{1}{2} B^{l}({\bf k^{\prime}})
B_{l}({\bf k}-{\bf k^{\prime}})] e^{i {\bf k}\cdot{\bf x}}  \label{aforce}
\eea
Comparing Eq. (\ref{fourier}) and Eq. (\ref{aforce}), 
we can find the relations
\bea
A+B= \frac{1}{2}   \nonumber \\
F(k)= \int d^{3}k^{\prime} B^{l}({\bf k^{\prime}})B_{l}({\bf k}
-{\bf k^{\prime}}) 
\eea

Then Eq. (\ref{atrace}) is written  by 
\bea
\int d^{3}k (\frac{1}{3} \delta^{i}_{j} - \frac{k^{i}k_{j}}{k^2})k^2 (\Phi({\bf k})+\Psi({\bf k}))
e^{i{\bf k}\cdot{\bf x}} = - 8\pi G a^{2} \int d^{3}k (\frac{1}{3} 
\delta^{i}_{j} - A \delta^{i}_{j} - B \frac{k^{i}k_{j}}{k^2}) F(k)
 e^{i{\bf k}\cdot{\bf x}}
\eea
Further we only treat the scalar density 
perturbations and traceless component of 
pressure perturbations, so  the right hand 
side should be proportional to 
$(\frac{1}{3} \delta^{i}_{j} - \frac{k^{i}k_{j}}{k^2})$.
Then we find that $A$ must have  value  of $\frac{1}{4}$.
Finally, we can write the traceless part for a given mode $k$,
\bea
k^2(\Phi +\Psi)= - 8 \pi G a^2 (\frac{1}{4} F(k))
\eea

\section{Spin-s Harmonics}
In this appendix we summarize the definition of spin-s function
and the property of spin-s harmonics. We mainly refer to
Ref.'s \cite{zaldarriaga} and \cite{hu}. 

A function $_sf(\theta, \phi)$ defined on the sphere is said
to have spin s if under a right-handed rotation of $(\hat{\bf e_1},
\hat{\bf e_2})$ by an angle $\psi$ it transforms as $_sf^{\prime}(
\theta,\phi)=e^{-is\psi}_sf(\theta,\phi)$. A spin-s functions
can be expanded in spin-s spherical harmonics, $_sY_{lm}
(\theta,\phi)$, which form a complete and orthonormal basis.
The spin-s harmonics are expressed as
\bea
 _sY_{lm}(\theta,\phi)= && e^{im\phi} \left[
\frac{(l+m)!(l-m)!}{(l+s)!(l-s)!}\frac{2l+1}{4\pi}\right]^{1/2}
{\rm sin}^{2l}(\theta/2) \sum_{r}
 \left( 
\begin{array}{c}
l-s \\ r 
\end{array} \right) 
\left( \begin{array}{c}
l+s \\ r+s-m  \end{array} \right) \nonumber \\
&& \times (-1)^{l-r-s+m}{\rm cot}^{2r+s-m}(\theta/2).
\eea
These set of functions satisfy the conjugatation, 
completeness and orthogonality
relations:
\bea
&& _sY_{lm}^{*}=(-l)^{m+s}_{-s}Y_{l-m}  \\
&& \int_{0}^{2\pi} d\phi \int_{-1}^{1} {\rm dcos} \theta 
_sY_{l^{\prime}m^{\prime}}^{*}(\theta,\phi)
_sY_{lm}(\theta,\phi)= \delta_{l^{\prime}l}\delta_{m^{\prime}m}, \\
&& {\sum_{lm}} _sY_{lm}^{*}(\theta,\phi)_sY_{lm}(\theta^{\prime},\phi^{\prime})
=\delta(\phi-\phi^{\prime})\delta({\rm cos}\theta- {\rm cos} \theta^{\prime}).
\eea
Finally the harmonics are related  to the ordinary spherical harmonics as
\bea
_{\pm}Y_{lm}= \left[\frac{(l-2)!}{(l+2)!}\right]^{1/2}
\left[\partial_{\theta}^{2}-{\rm cot} \theta \partial_{\theta}
\pm \frac{2i}{{\rm sin} \theta}(\partial_{\theta}
-{\rm cot} \theta)\partial_{\phi}
-\frac{1}{{\rm sin}^{2}\theta}\partial_{\phi}^{2} \right] Y_{lm}.
\eea

\newpage
\begin{figure}[htbp]
\centerline{\epsfig{file=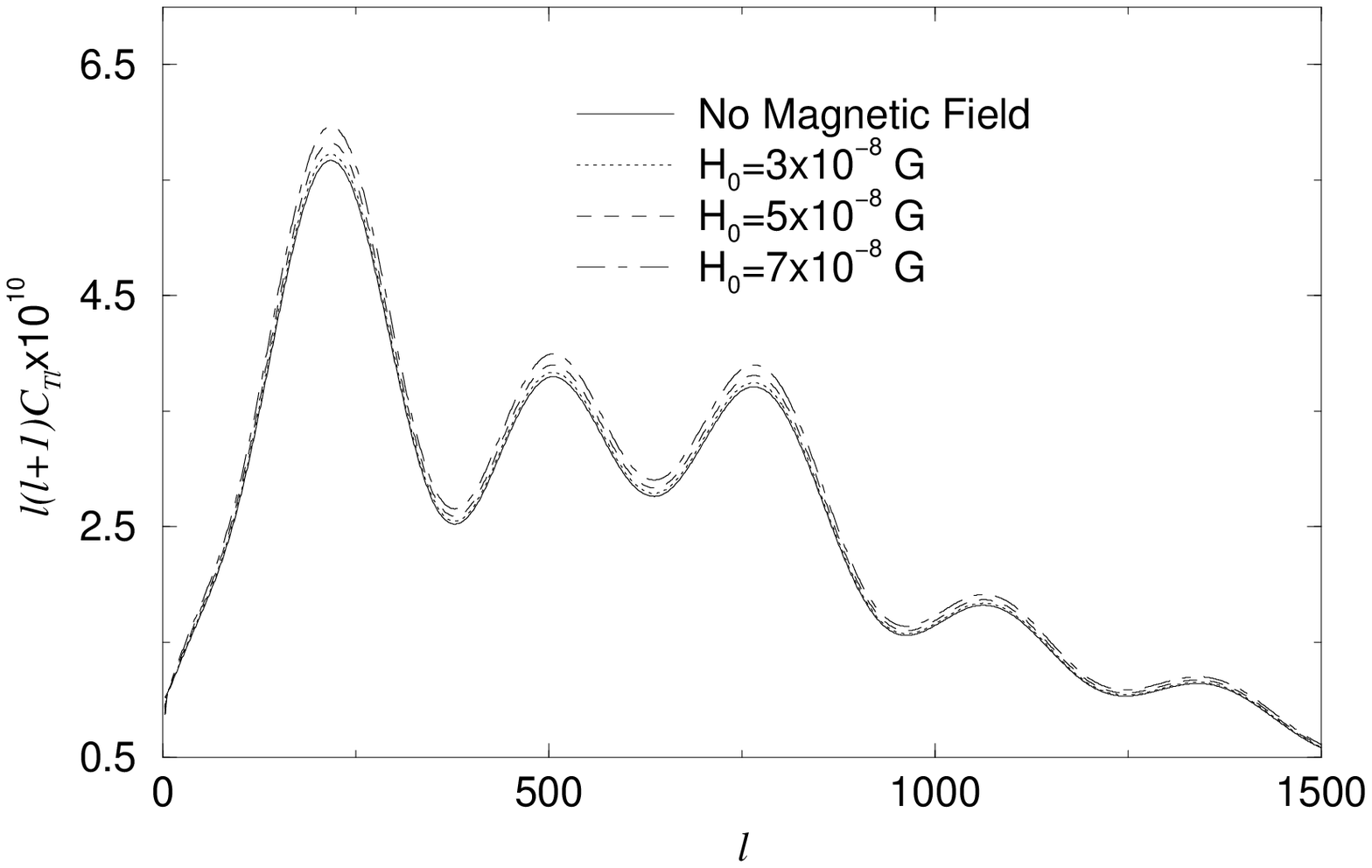,width=0.9\linewidth,height=0.4\textheight}}
\caption{The angular power spectrum of temperature fluctuations with 
the magnetic field strength  $B_{\lambda}=3\times 10^{-8}G, 
5\times 10^{-8}G, 7\times
10^{-8}G$ with spectral index $q=1$ for $\lambda=0.1h^{-1}
{\rm Mpc}$.}

\centerline{\epsfig{file=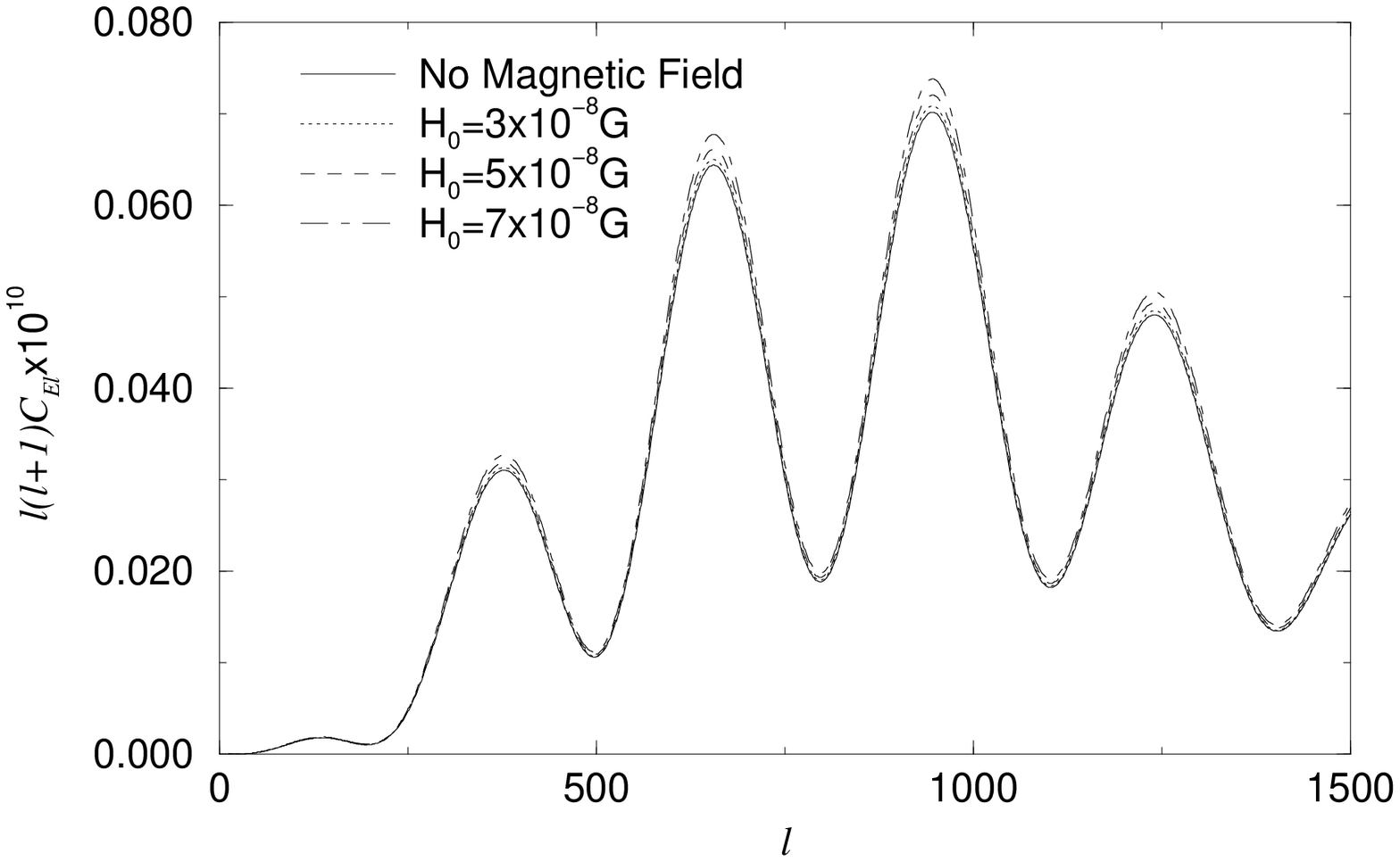,width=0.9\linewidth,height=0.4\textheight}}
\caption{The $E$ mode polarization spectrum $l(l+1)C_{El}$  for $q=1$.}

\epsfig{file=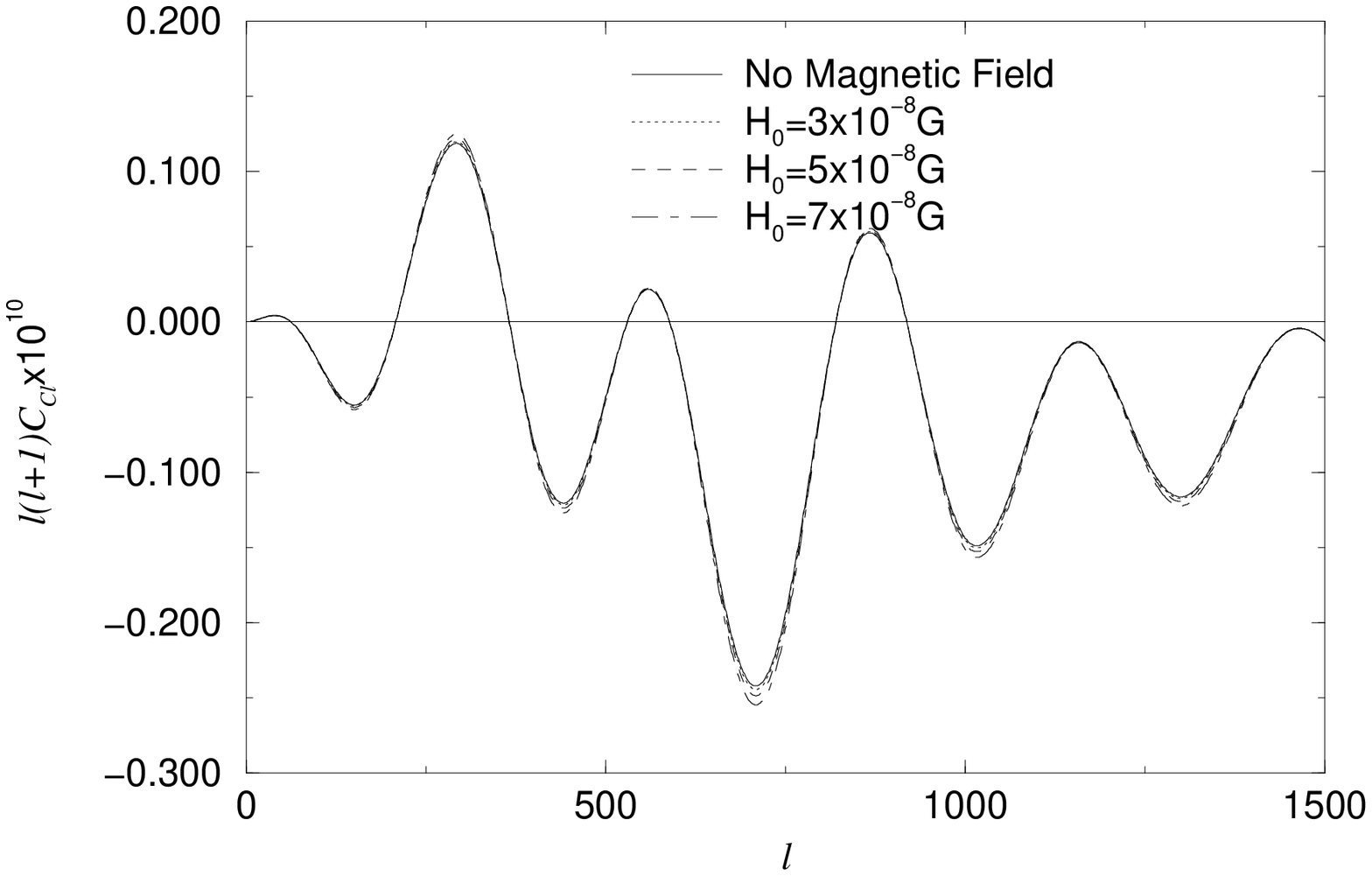,width=0.9\linewidth,height=0.4\textheight}
\caption{Temperature and polarization cross-correlations, $l(l+1)C_{cl}$,
 for $q=1$.}

\epsfig{file=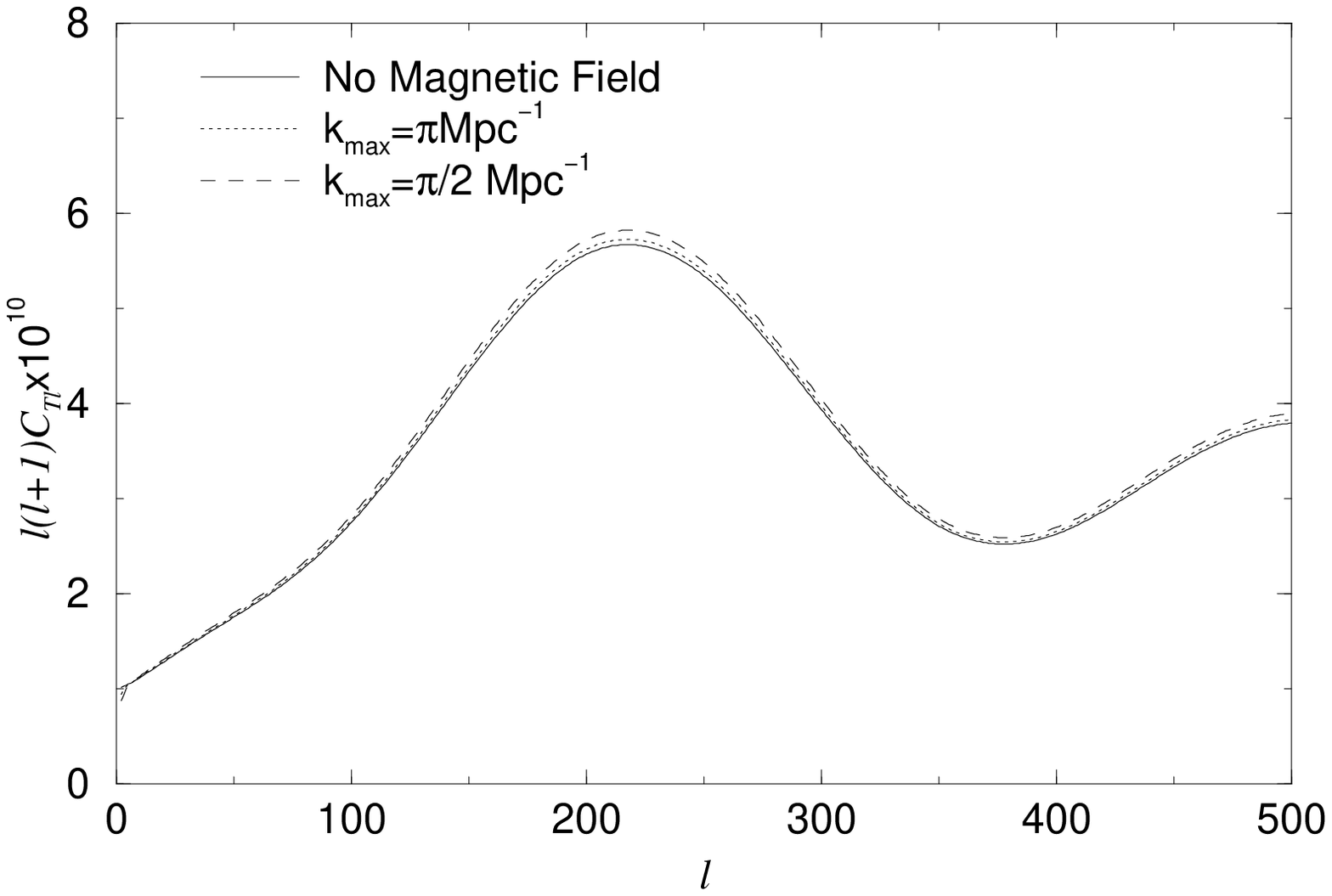, width=0.9\linewidth,height=0.4\textheight}
\caption{The angular power spectrum of temperature fluctuations with
the value of the magnitude of 
cut-off wavevector $k_{max}=\pi Mpc^{-1},\pi/2 Mpc^{-1}$
for $B_{\lambda}=5\times 10^{-8}G$ for $\lambda=0.1h^{-1}{\rm Mpc}$.}

\epsfig{file=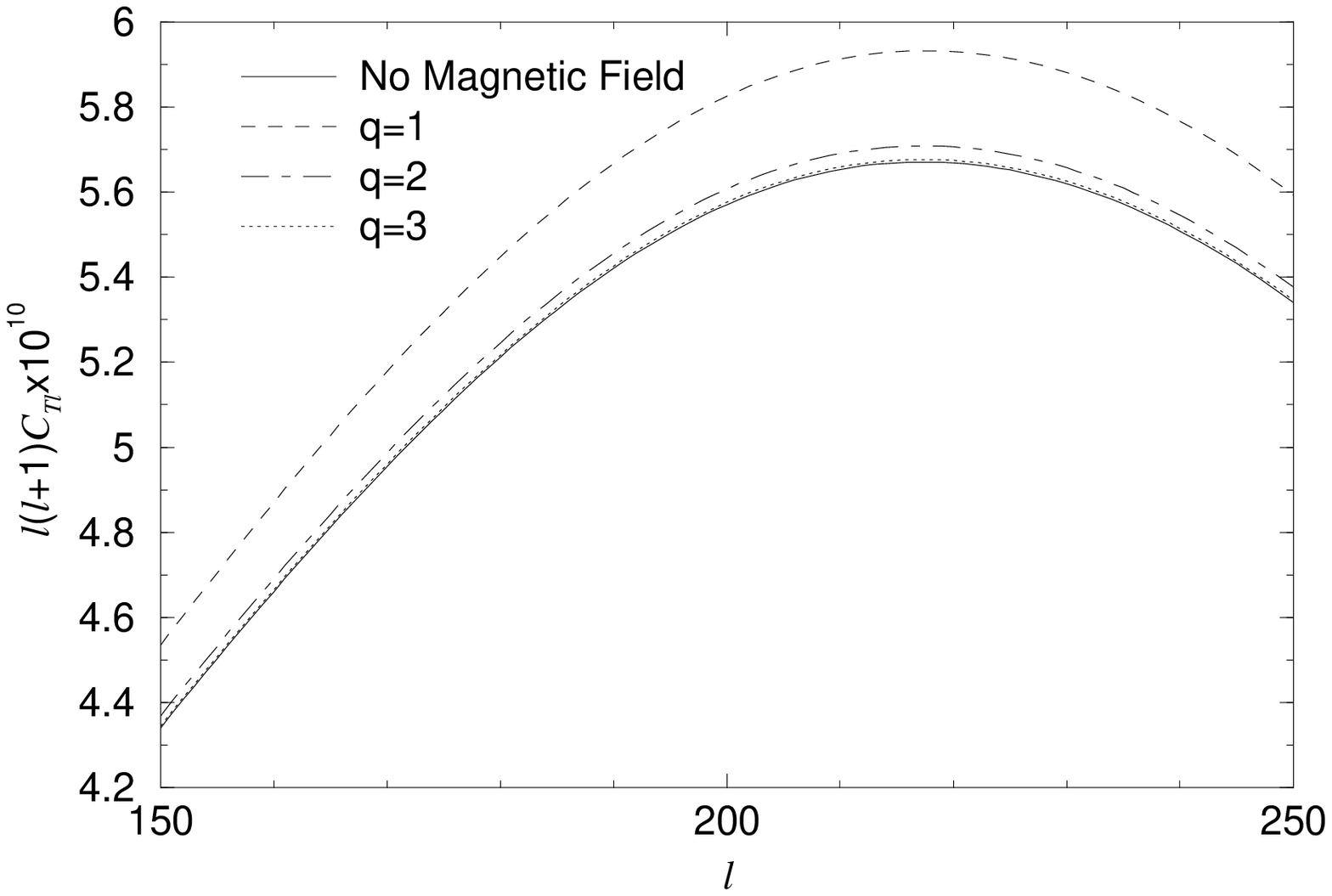, width=0.9\linewidth,height=0.4\textheight}
\caption{Temperature anisotropy with the spectral index of magnetic
field energy density $q=1,2,3$ for $B_{\lambda}=5\times 10^{-8}G$
for $\lambda=0.1h^{-1} {\rm Mpc}$.}

\end{figure}
\end{document}